\documentclass[a4paper,12pt]{article}
\usepackage[utf8]{inputenc}
\usepackage{graphicx}
\usepackage{booktabs}
\usepackage{natbib}
\usepackage{amsmath}
\usepackage[top=2in, bottom=1.5in, left=3cm, right=3cm]{geometry}

\newcommand{\Ham}{\mathcal{H}}

\newcommand{\piclength}{0.45\linewidth}
\title{Measuring switching processes in financial markets with the 
Mean-Variance spin glass approach}
\author{J. Jurczyk}

\begin{document}

\maketitle

\begin{abstract}
In this article we use the Mean-Variance Model in order to measure the current 
market state. In our study we take the approach of detecting the overall 
alignment of portfolios in the spin picture. The projection to the ground-states
enables us to use physical observables in order to describe the current state of
the explored market. The defined magnetization of portfolios shows cursor 
effects, which we use 
to detect turmoils.
\end{abstract}

\section{Introduction}
There are different methods in order to estimate switching in market 
behavior. It reaches from Markov switching processes 
\cite{hamilton1989new,schaller1997regime}
to newer methods where physical phase transitions are used to describe a 
switching process\cite{preis2011switching} as well as utilizing search engine 
queries\cite{preis2013quantifying,preis2010complex,kristoufek2013can,
kristoufek2013bitcoin,moat2013quantifying}. 
These techniques rely on the assumption, that market moves or social transitions 
can be detected by observing the behavior of online queries. These queries 
represent precursor effects for phase transitions.
All approaches have a similar goal: Identifying risks within the economy.

Our aim in this article is to identify switching processes within the 
Mean-Variance Model introduced by Markowitz\cite{markowitz1952portfolio}. We 
use the portfolios of the efficient frontier under the assumption that these 
hold important information about the current market state.

We demonstrate that the optimal portfolios, in the magnetization picture 
with free spins aligning to the overall external market field, detect 
financial turmoils. This measure shows a not decaying risk of market 
switching since the 
Euro crisis. This can be seen in other studies containing the systemic 
risk\cite{billio2012econometric,jurczyk2015comparing}.

\section{The Mean-Variance Model}
We consider the Mean-Variance Model, introduced by Markowitz in 1952 
\cite{markowitz1952portfolio}. The objective is to maximize
the average return 
\begin{equation}
\overline{\mathcal{R}}\left(\mathbf{P}\right)=\sum\limits_{i=1}^{N}R_{i}\sigma_{
i}
\end{equation}
and minimize the risk defined by
\begin{equation}
 \mathrm{var}\left(\mathbf{P}\right)=\sum\limits_{i,j}\sigma_{i}C_{ij}\sigma_{j}
\end{equation}
where $\mathbf{P}$ is the portfolio vector for a specific historical time 
window. $\sigma_{i}$ is the weight of an
asset $i$, $R_{i}$ is the average return of asset $i$ for that time window, 
$C_{ij}$ is the corresponding covariance
matrix and $N$ the number of considered assets.  Since an investor has limited 
resources, the weights need to fulfill
the constraint
\begin{equation}
 \sum\limits_{i}^{N}\left|\sigma_{i}\right|\stackrel{!}{=}1
 \label{eq:constraint}
\end{equation}
Therefore $\sigma_{i}$ can range from $\left[-1,1\right]$ and short-selling is 
allowed. An investor can sell assets
which he does not own and rebuy later for a lower price to benefit from a price 
drop.  This gives us the analogy
to spins, which can point in any direction and projected on an arbitrary 
direction which can be seen as a magnet model 
\cite{rosenow2002portfolio}. As Jahan , our objective 
function is regarded as a physical Hamiltonian
\begin{equation} 
\Ham_{\lambda}\left(\mathbf{P}\right)=-\lambda\cdot\overline{\mathcal{R}}
\left(\mathbf{P}
\right)+(1-\lambda)\cdot\mathrm { var } \left(\mathbf { P } \right)
\label{eq:ham}
\end{equation}
where $\lambda\in[0,1]$ is responsible for balancing the two 
objectives\cite{jahan2010local}. 
\subsection{Identifying the switching process}
For each $\lambda\in[0,1]$ an optimal portfolio $\mathbf{P}_{0}$ exists, which 
balances the risk and return, resulting in an efficient frontier. Each optimal 
portfolio $\mathbf{P}_{0}$ on the efficient 
frontier has a ``magnetization'' value $m=\sum_{i}\sigma_{i}$, which is a risk 
indicator, since the magnetization depends on the weighing parameter $\lambda$ 
from Eq. (\ref{eq:ham}) at each each measurement at time $t$.

The magnetization $m$ can range from all assets are sold-short
($m=-1$) to a pure buying suggestion ($m=1$). These two market stages are called
bearish or bullish market.

In a bullish market the magnetization $m$ ranges from approximately zero for
$\lambda=0$ to positive values $m>0$. The opposite is the case for a bearish 
market. 

\begin{figure}[ht]
\begin{tabular}{cc}
 \resizebox{\piclength}{!}{\includegraphics{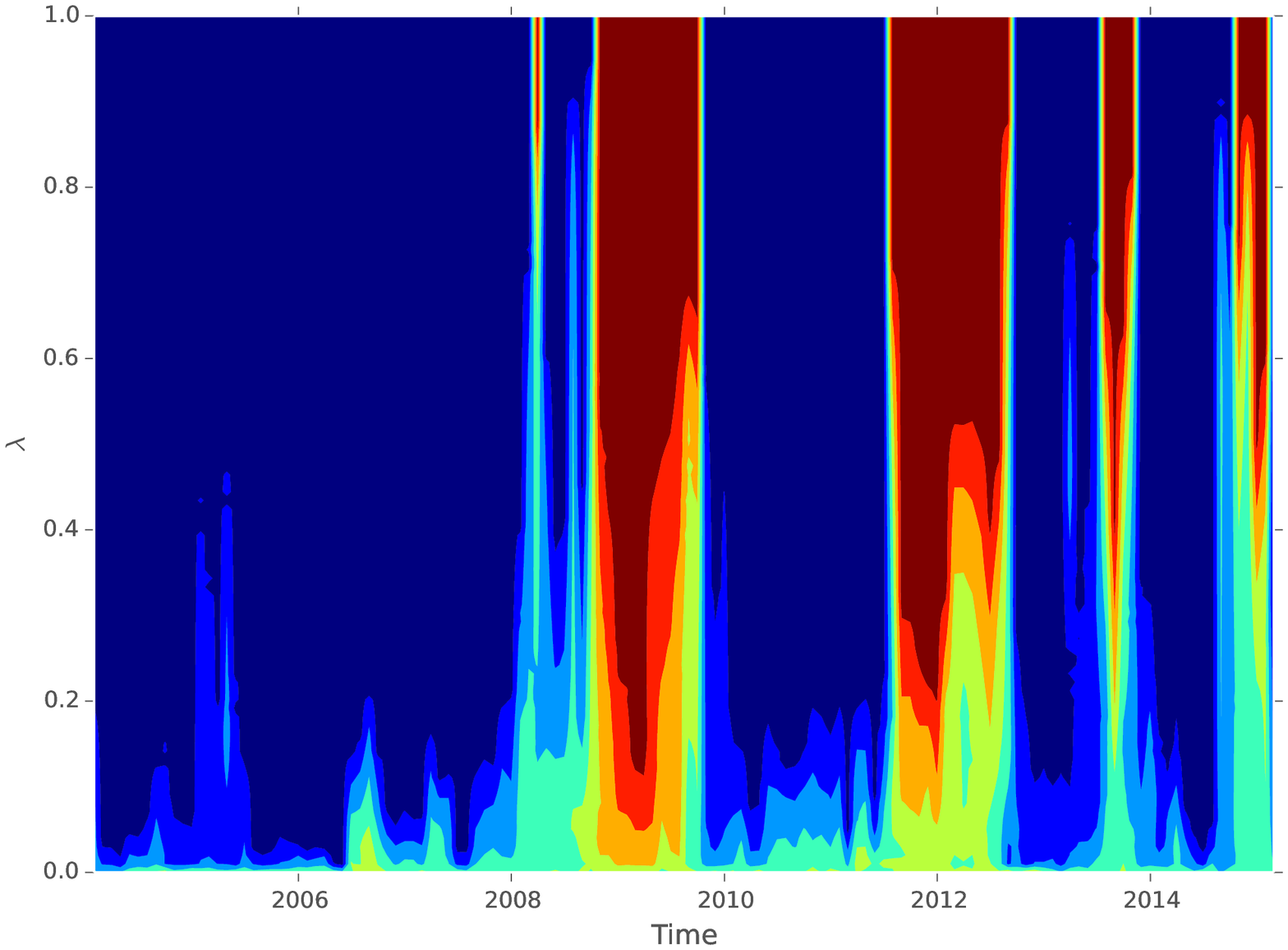}} &
 \resizebox{\piclength}{!}{\includegraphics{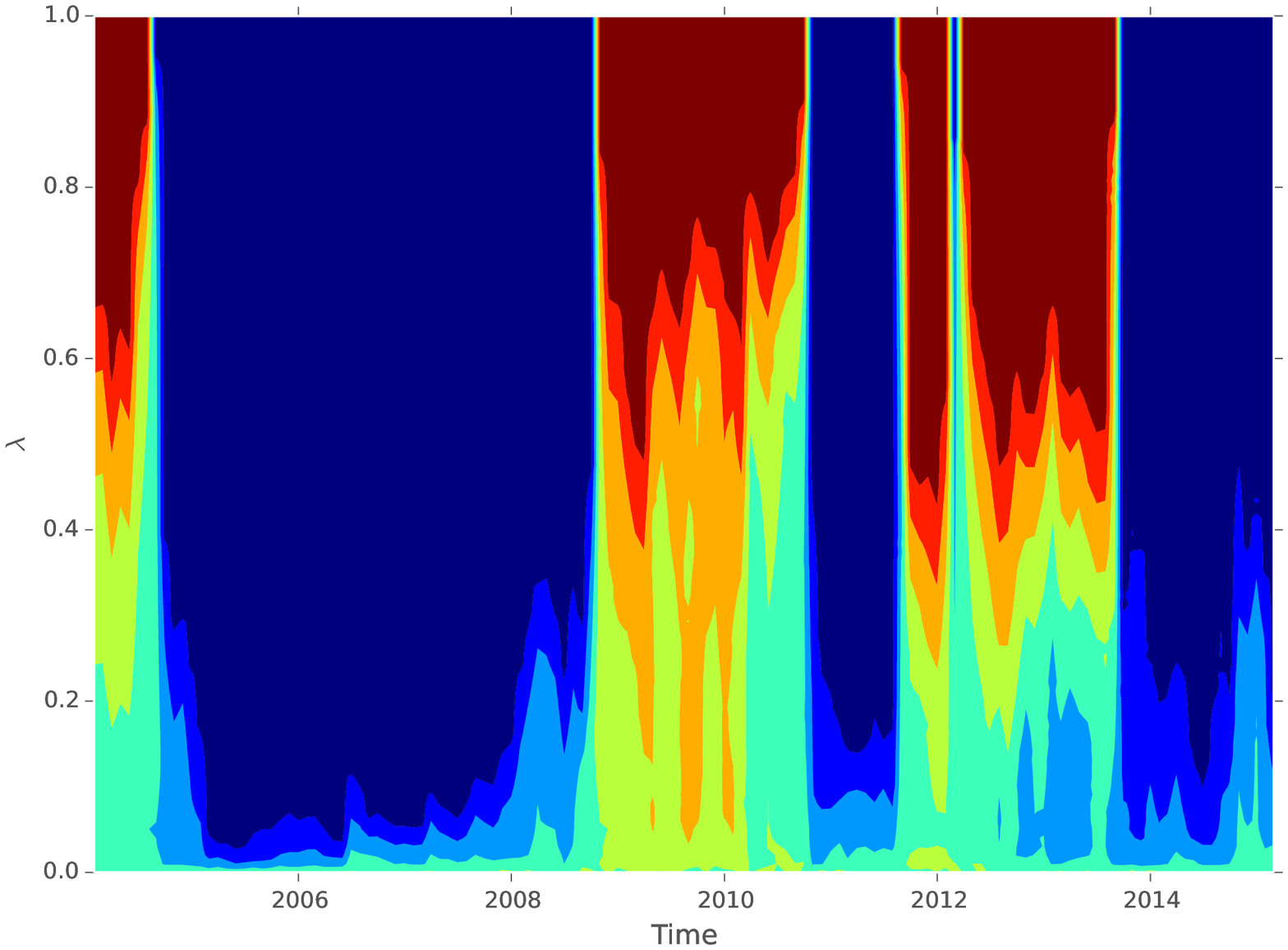}} \\
\end{tabular}
\caption{On the left the data for the Mean-Variance Model has a time-frame of 
one year. One right the time-frame was set to two years. Both measurements rely 
on monthly returns of the assets, which belong to the DAX 30. The financial 
crisis of 2008 has a build up starting in September 2007, while the Euro crisis 
suddenly changes. (blue corresponds to $m>0$, red $m<0$)}
 \label{fig:timevsmag}
\end{figure}

In Figure \ref{fig:timevsmag} we 
used the assets of the German index DAX 30 in order to measure the market 
state. The 
magnetization $m$ clearly shows the two main crisis of the last ten years. The 
financial crisis of 2008 and the Euro crisis of 2011. 
With increasing $\lambda$ the magnetization $m$ starts to decrease in September 
2007, which reflects the uncertainty of a Mean-Variance portfolio of the market 
direction. When the Lehman bankruptcy news hit the market, the magnetization 
finally flips in September 2008 to a bearish phase.

The second major crisis started in 2010, when several Euro states were unable 
to place national bonds. Until September 2011 the market was not concerned but 
then suddenly switched to a negative trend. This was due to the fact that in 
this time a fear of a Euro crash was amongst many investors.

\begin{figure}[ht]
\begin{tabular}{cc}
\resizebox{\piclength}{!}{\includegraphics{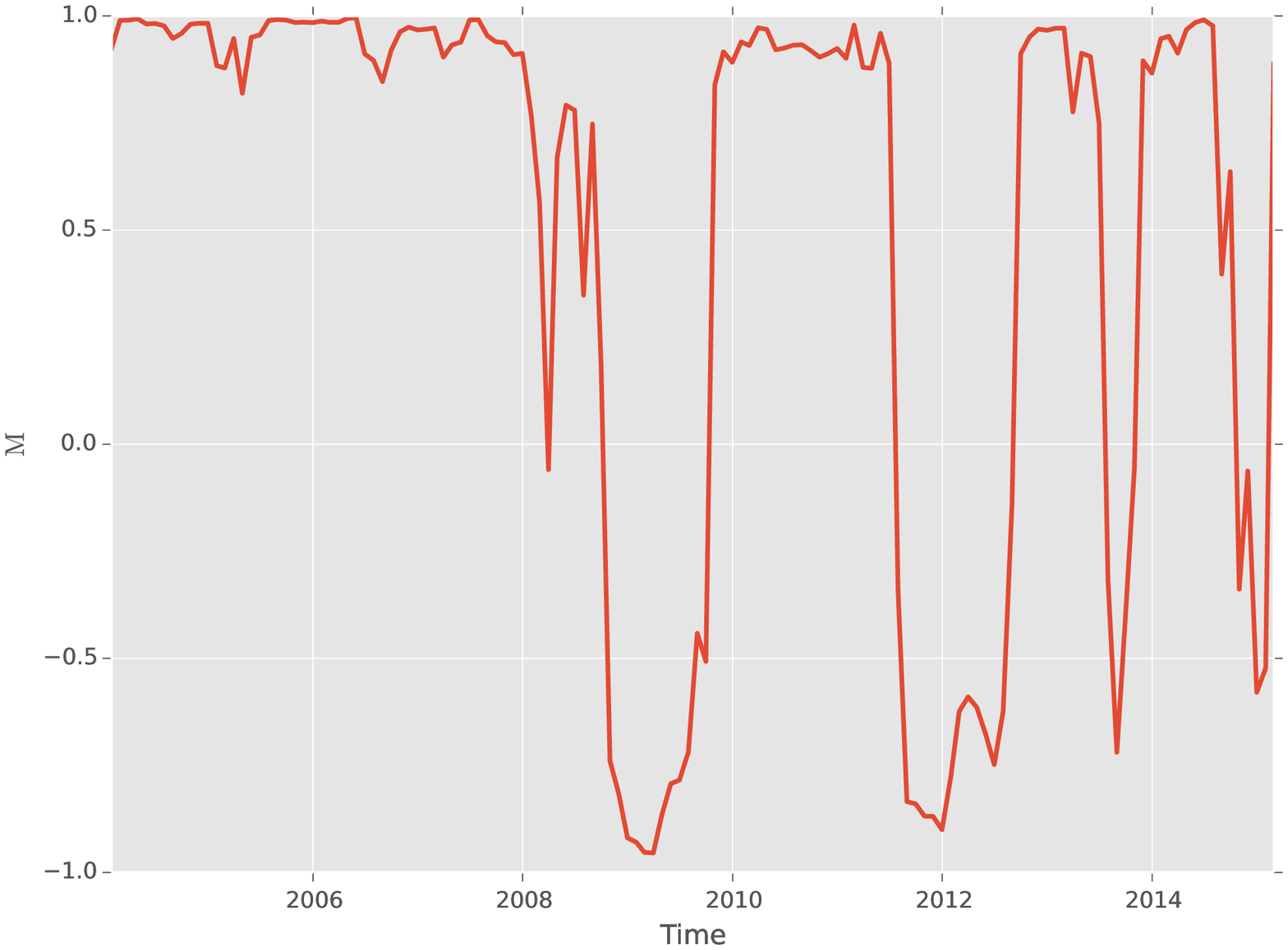}} &
\resizebox{\piclength}{!}{\includegraphics{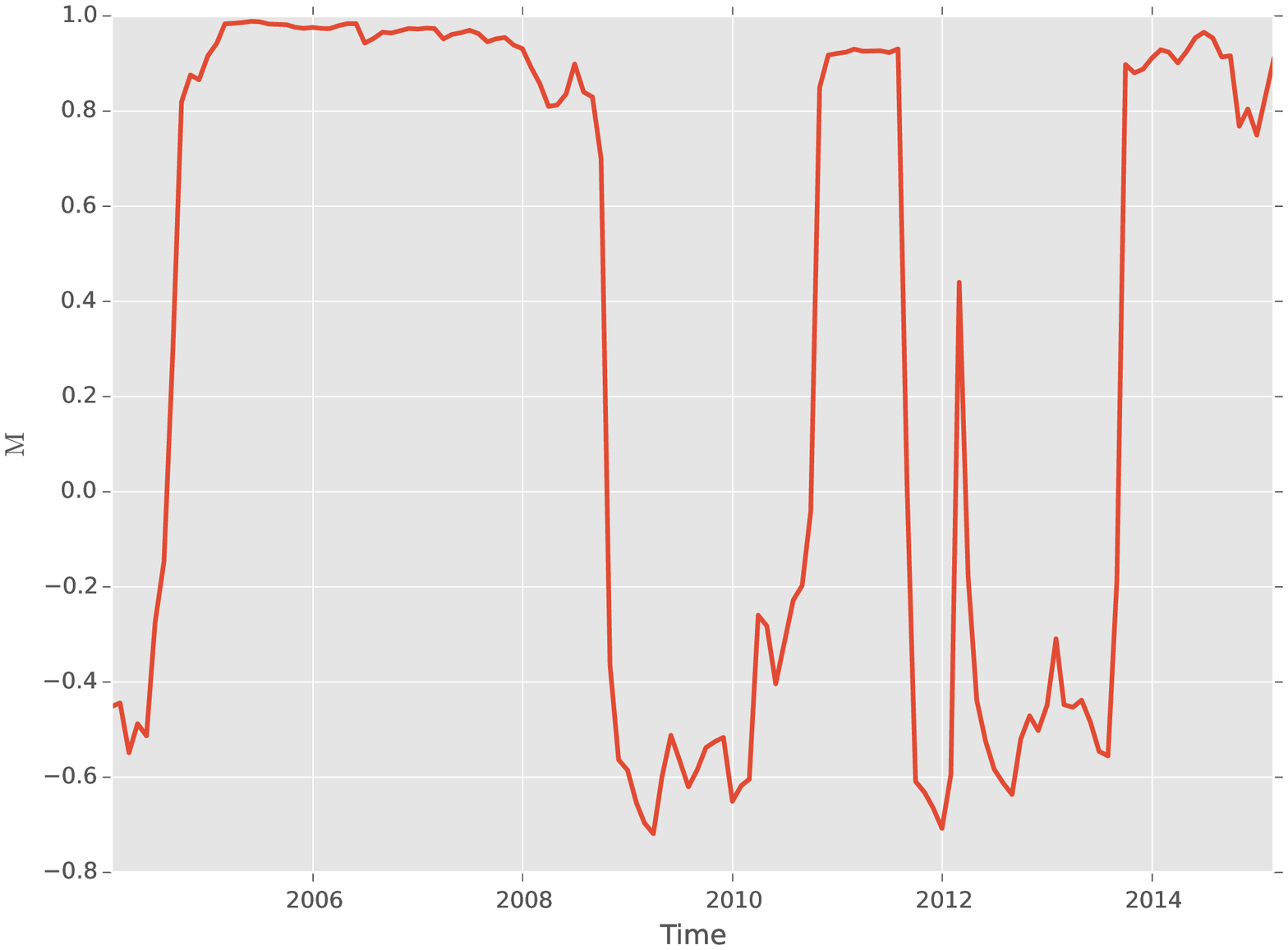}} \\
\end{tabular}
\caption{On the left the data for the Mean-Variance Model has a time-frame of 
one year. One right the time-frame was set to two years. Both measurements rely 
on monthly returns of the assets, which belong to the DAX 30.}
 \label{fig:timevsimag}
\end{figure}

The integral 
\begin{equation}
 \mathrm{M}=\int\limits_{0}^{1}\mathrm{d}\lambda \cdot m\left(\lambda\right)
\end{equation}
is proportional to the tendency of the optimal portfolios to be bearish or bullish.
The value of $\mathrm{M}$ ranges between $-1$ and $1$. While these are the upper bounds 
and real data are not achieved since for small values of $\lambda$ a 
distribution of assets is 
wanted and leads to $m$ close to $0$. 
In figure \ref{fig:timevsimag} we show the $\mathrm{M}$ in respect to the time.

In both figures \ref{fig:timevsimag} and \ref{fig:timevsmag} the stability of 
the ground-states starts to
decay in 2008. This shows that the optimal portfolios of the efficient frontier 
become aware of the risk in the market 
and can only compensate that risk by shorting certain assets.

\subsection{Cursor effects in the switching process}
\begin{figure}[htbp]
\begin{tabular}{cc}
 \resizebox{\piclength}{!}{\includegraphics{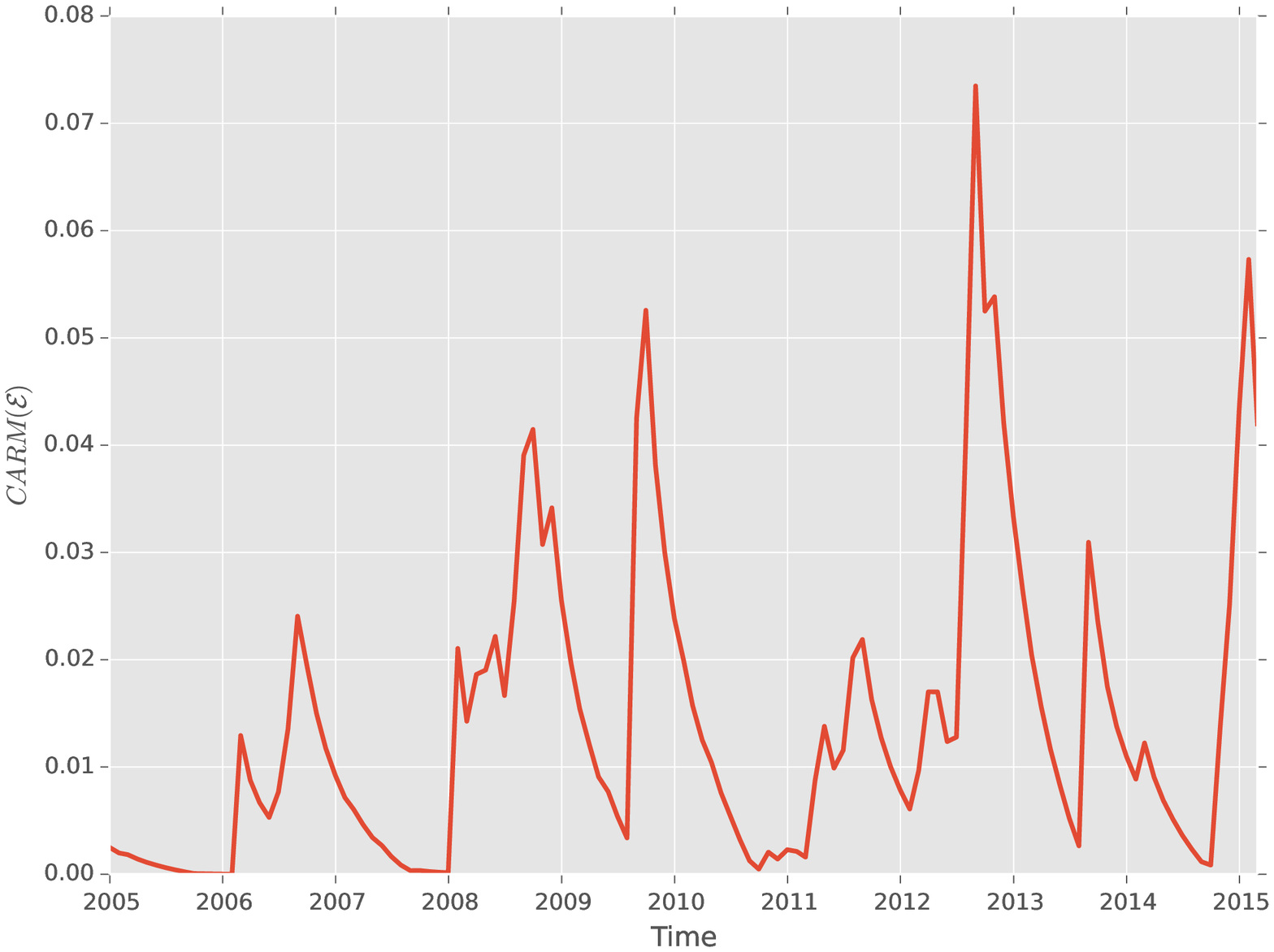}} &
 \resizebox{\piclength}{!}{\includegraphics{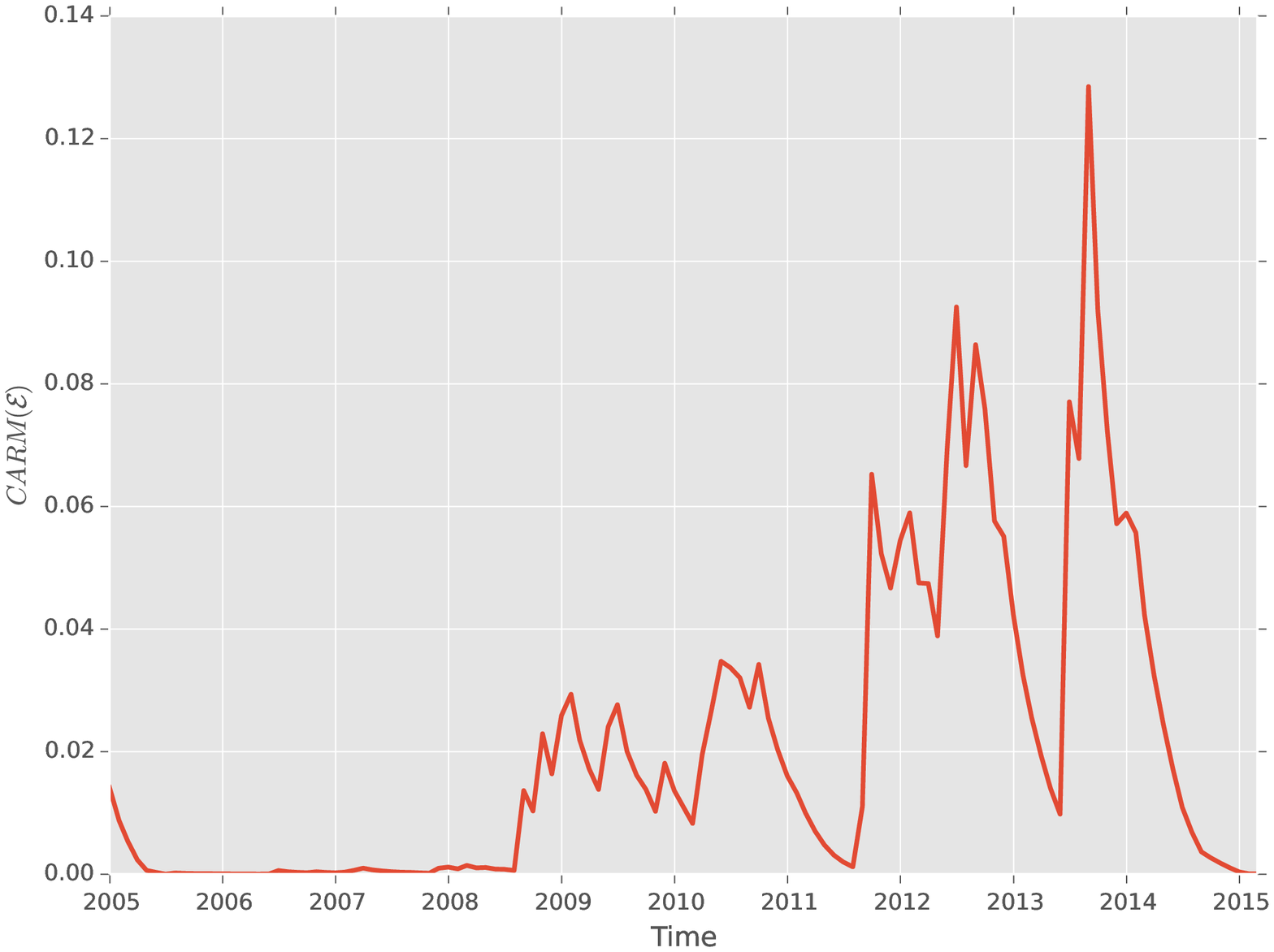}} \\
\end{tabular}
\caption{(left) $CARM$ of the event in Eq. \ref{eq:event_zero} with a 
time window of one year. (Right) A time window of two years is depicted. Note that since the beginning 
of the Euro crisis in both time frames the $CARM$ does not settle down and keep 
surging.The measurements used monthly returns.}
\label{fig:event_zero}
\end{figure}

\begin{figure}[htbp]
\begin{tabular}{cc}
 \resizebox{\piclength}{!}{\includegraphics{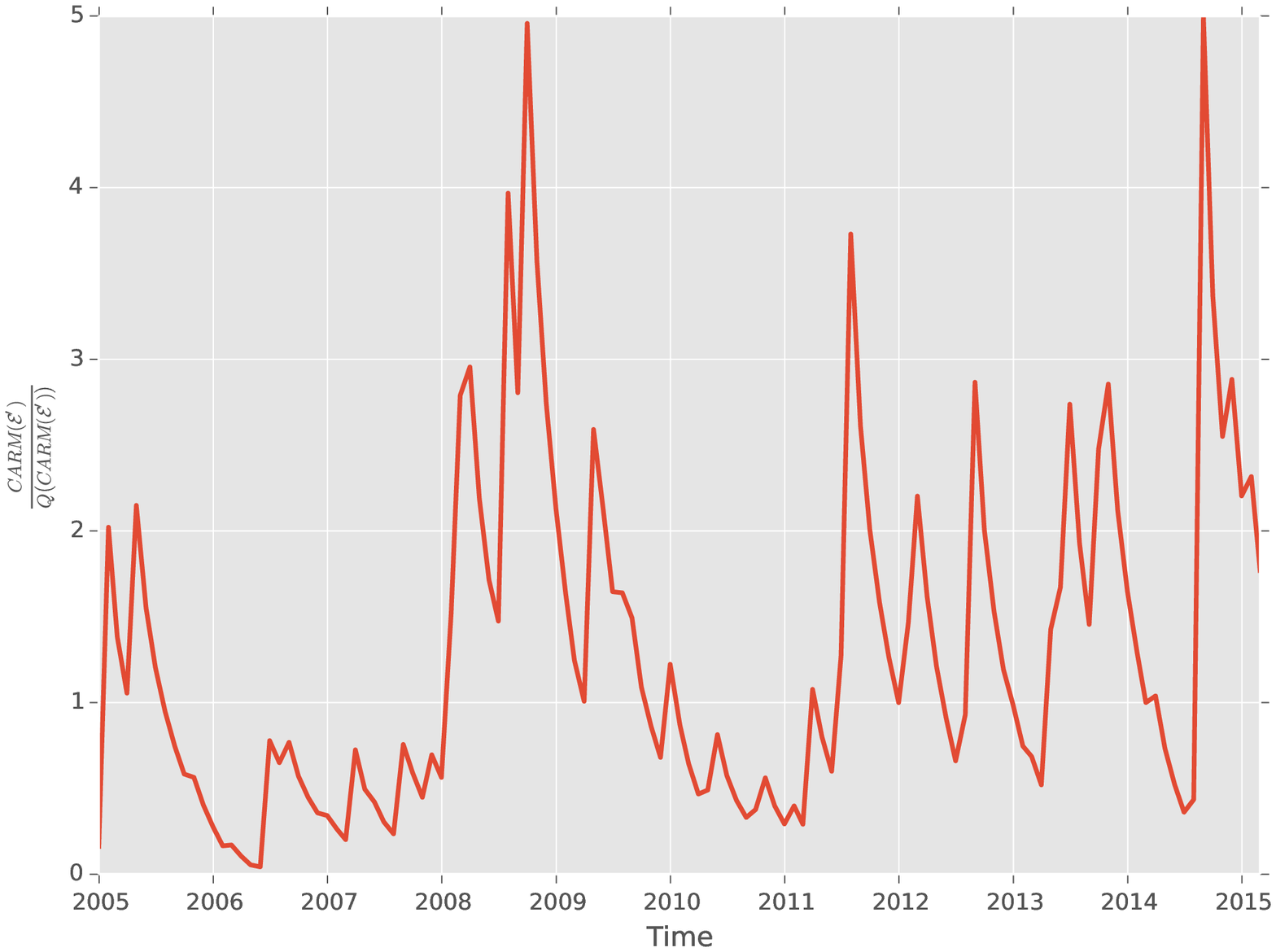}} &
 \resizebox{\piclength}{!}{\includegraphics{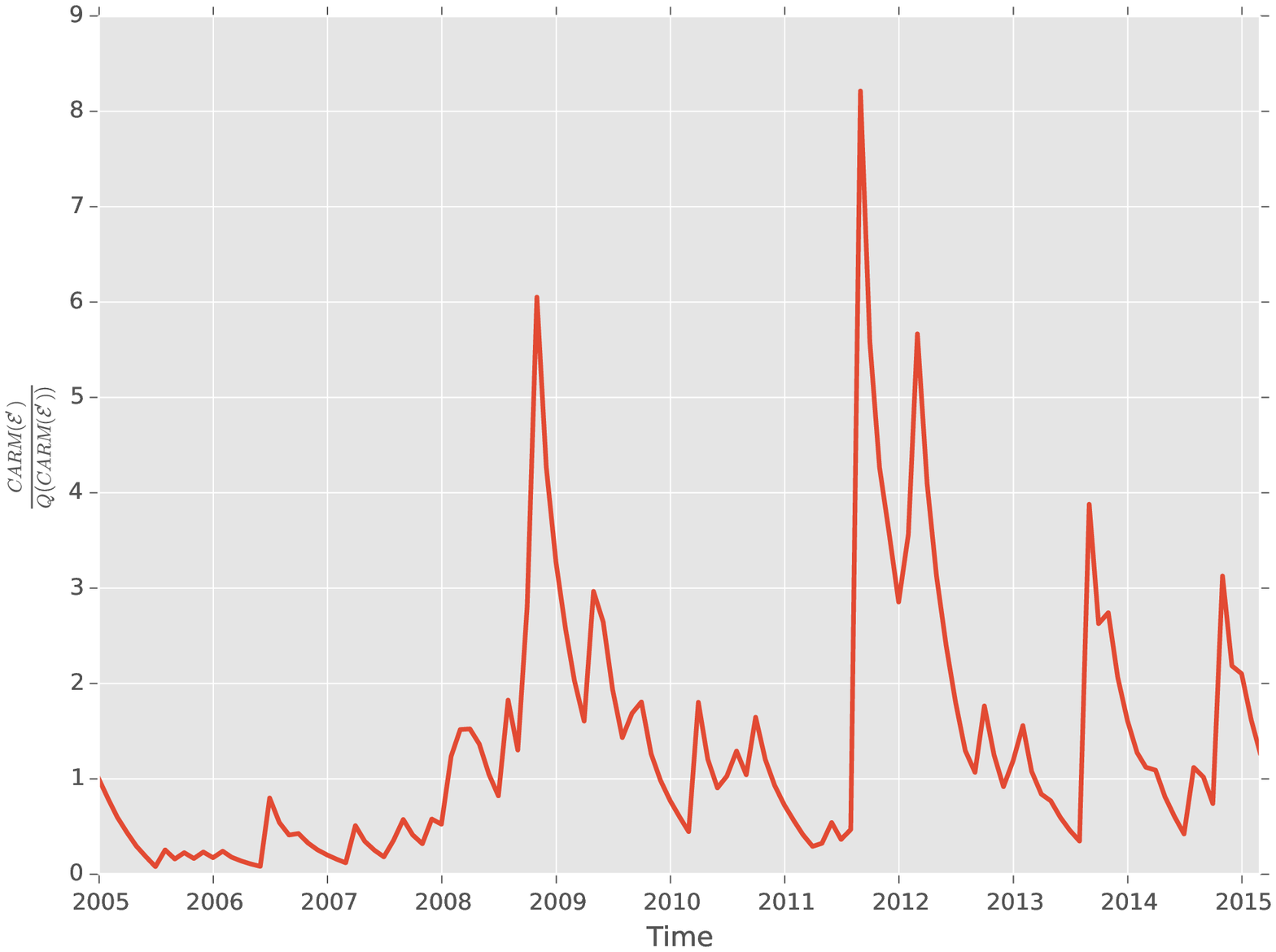}} \\
\end{tabular}
\caption{The event $\mathcal{E}$ is defined by the smallest $\lambda$ before 
$(\lambda)$ reaches its absolute maximum. (left) $CARM$ with a time 
window of one year. (Right) $CARM$ with a time window of two years. Notice that the CARM was 
normalized by the median $Q$}
\label{fig:event_phase}
\end{figure}

The uncertainty within the dataset is represented by the magnetization as 
described above. We observed two effects within the magnetization before the 
two crisis.
The first is that we noticed cumulated events of the form
\begin{equation}
 \mathcal{E}=\begin{cases}
              \lambda_{0}^{crit}&\exists\lambda_{0}^{crit} | 
m(\lambda_{0}^{crit} )=0\\
              0&\text{else}\\
             \end{cases}
\label{eq:event_zero}
\end{equation}
which translates to: Although there is emphasize on the portfolio to generate 
return, the discussion to invest in a bearish or bullish trend is not given. 
Hence the magnetization $m=0$.
We also measured the smallest $\lambda$, where $m(\lambda)$ reaches its 
absolute maximum(Event $\mathcal{E}'$). 
In figure \ref{fig:event_zero} and \ref{fig:event_phase}, we show that these 
two events accumulate before the crisis in 2008 and 2011. The upbuilding process 
can be demonstrated by a cumulative averaged rolling mean 
(CARM).\cite{jurczyk2014timedependent}
\begin{equation}
 CARM(t)=\frac{1}{N}\sum_{i=1}^{N}\sum_{j=0}^{i}\frac{\mathcal{E}(t-j\cdot\delta)}{i}
\end{equation}
where $N\cdot\delta$ is the maximum time-window one considers of the event $\mathcal{E}$ 
at time $t$.

In both events one can identify the two crisis. The rise in the 
$CARM(\mathcal{E})$ takes place in January 2008($\Delta=\textrm{1y}$) 
and June 2008($\Delta=\textrm{2y}$). This references to a higher risk within 
the DAX 30 at that time. The peak in September 2008 finishes the upbuilding 
switching process (see figure \ref{fig:timevsmag}). 

A similar structure can be observed with the $CARM(\mathcal{E}')$. A surge 
occurs in January 2008. The Lehman crash in September 2008 leads to the 
distinct 
peak of the crisis. 

The Euro crisis has a different from. A first switch from bullish to bearish 
took place in September 2011. But both events show distinct spires since 2011. 
This circumstance leads to the conjecture, that the portfolios chosen by the 
Mean-Variance Model do not settle for a phase.

\section{Conclusion}
We showed that the Mean-Variance Model with a short-selling option can be used 
to measure the current state of an underlying market by utilizing the 
portfolios of the efficient frontier as a spin grid under the constraints of 
risk and returns.
The are three event types, which are proportional to the minimal risk within the
underlying market, the integral over the magnetization and the two cursor 
events.
Therefore we connected the overall market state to the ground-states of the 
portfolio 
given by the efficient frontier of the Mean-Variance Model.
By introducing a memory measure $CARM$ we detect upbuilding
risk bubbles.
\bibliographystyle{plainnat}
\bibliography{mean_variance}
\end{document}